# Direct imaging of current-induced antiferromagnetic switching revealing a pure thermomagnetoelastic switching mechanism

Authors:

H. Meer[1], F. Schreiber[1], C. Schmitt[1], R. Ramos[2], E. Saitoh[2,3,4,5,6], O. Gomonay[1], J. Sinova[1,7], L. Baldrati[1*], and M. Kläui[1,8*]

[1]Institute of Physics, Johannes Gutenberg-University Mainz, 55128 Mainz, Germany

[2]WPI-Advanced Institute for Materials Research, Tohoku University, Sendai 980-8577, Japan

[3]Institute for Materials Research, Tohoku University, Sendai 980-8577, Japan

[4]Advanced Science Research Center, Japan Atomic Energy Agency, Tokai 319-1195, Japan

[5]Center for Spintronics Research Network, Tohoku University, Sendai 980-8577, Japan

[6]Department of Applied Physics, The University of Tokyo, Tokyo 113-8656, Japan

[7]Institut of Physics, Academy of Sciences of the Czech Republic, Praha 11720, Czech Republic

[8]Graduate School of Excellence Materials Science in Mainz, 55128 Mainz, Germany

*Electronic Mail: lbaldrat@uni-mainz.de, Klaeui@Uni-Mainz.de

**Abstract:**

We unravel the origin of current-induced magnetic switching of insulating antiferromagnet/heavy metal systems. We utilize concurrent transport and magneto-optical measurements to image the switching of antiferromagnetic domains in specially engineered devices of NiO/Pt bilayers. Different electrical pulsing and device geometries reveal different final states of the switching with respect to the current direction. We can explain these through simulations of the temperature induced strain and we identify the thermomagnetoelastic switching mechanism combined with thermal excitations as the origin, in which the final state is defined by the strain distributions and heat is required to switch the antiferromagnetic domains. We show that such a potentially very versatile non-contact mechanism can explain the previously reported contradicting observations of the switching final state, which were attributed to spin-orbit torque mechanisms.

**Introduction:**

Spintronic devices to date rely heavily on ferromagnets (FM) as active elements to store and manipulate information. Antiferromagnets (AFMs) are prime candidates to replace FMs as active elements in future spintronic devices since they potentially offer a higher bit packing density, resulting from the absence of magnetic stray fields, stronger resilience to external fields[1] and ultrafast switching speeds[2]. In particular, the investigation of insulating AFMs has recently sparked interest, because their low magnetic damping enables the transport of spin currents over long distances[3], potentially serving as a basis for low power devices. However, the absence of a net magnetic moment makes AFMs more difficult to manipulate compared to FMs.

In view of applications, it is of paramount importance to be able to perform the read-out and write operations fast and efficiently and to understand the underlying switching mechanism.

In collinear antiferromagnets, magnetic information can be stored in the orientation of the Néel vector **n**. In AFM/heavy metal (HM) bilayers the orientation of the Néel order can be read electrically via the spin Hall magnetoresistance (SMR)[4,5]. However, some reports interpreted the electrically detected SMR signals after current pulses as mere results of electrical heating and electromigration of the platinum wire[6,7] or a combination with resistive switching and torques due to thermomagnetoelastic effects[8]. Imaging of the magnetic domains, however, allows to unambiguously determine the presence of magnetic switching[9,10].

In the metallic antiferromagnets CuMnAs[11] and $Mn_2Au$[12] magnetic switching, due to current pulses inducing spin torques, relies on the special crystal symmetries of these materials, leading to current-induced staggered bulk Néel spin orbit torques. In insulating AFMs the underlying switching mechanism is highly debated. So far, different final states of the Néel vector **n** with respect to the current direction **j** have been reported in different collinear AFMs (**n** ∥ **j**) for NiO[9], (**n** ⊥ **j**) for CoO[13] and for different device and pulsing geometries in NiO (**n** ∥ **j**[9,14] and **n** ⊥ **j**[15]), so that different switching mechanisms have been put forward as the origin. Most authors proposed mechanisms related to antidamping-like spin orbit torques (SOT) to explain the observed electrical switching in insulating AFM/HM bilayers: the torques were considered acting on the uncompensated moments focusing on one sublattice[15], as well as SOTs related to both sublattices[14] or acting on the domain walls[9].

To understand the origin of these different reports, primarily based on electrical readout, one needs to directly image the magnetic switching. NiO is an ideal material system to investigate the switching mechanism of insulating AFMs. Due to its high Néel temperature of 523 K in the bulk[16], its insulating collinear antiferromagnetic phase and the strong magnetoelastic coupling[17] that allows one to image the AFM domain structure with Kerr microscopy[10,18]. This lab-based imaging approach can be combined with concurrent electrical readout to reliably identify a magnetic switching of NiO/Pt bilayers[10].

Additionally, the samples need to allow one to be able to disentangle the effects of SOTs, directly generated by the current, and the thermomagnetoelastic torques, generated by the current-induced temperature gradients. This can be achieved by engineering the device geometry, e.g. one can generate temperature gradients in regions where no current is flowing. This is necessary due to the similar linear functional dependency of the effective fields from SOTs and thermomagnetoelastic effects on the current density $j$[13], while other heat induced torques depend quadratically on $j$[8,9,19].

In electrical writing schemes to date, high current densities are used to switch the AFM, resulting in electromigration and irreversible device degradation[7], which is impractical for device applications. Thus, it is of paramount importance to understand the underlying switching mechanism to engineer reliable, fast and efficient writing schemes for insulating AFMs.

Here, by combining electrical measurements, table-top direct imaging of the magnetic domains and simulations of the thermally induced strain in specially engineered device geometries, we find that the final state of the switching does not solely depend on the direction of $j$, but rather on the device geometry. We show that the switching is enabled by a combination of strain and heat, while the current flow through the device is not necessary to induce the switching. We explain these findings by a thermomagnetoelastic switching mechanism of the Néel order in NiO/Pt thin films. Finally, we discuss how this mechanism can explain the different switching final states previously reported for the NiO/Pt system, based on the different geometries used.

**Results:**

**Effect of the device geometry.** To investigate the contribution of current-induced SOTs, heat and thermomagnetoelastic effects, we first consider the AFM switching in different device geometries. Note that, according to the SOT mechanism, the final state of the switching depends only on the direction of the current *j*, while in the case of the thermomagnetoelastic switching the final state is determined by the distribution and orientation of the temperature gradients[8,13]. To disentangle the two mechanisms, we employ two devices, patterned with different orientation on the same sample, aiming to obtain orthogonal strain while using the same direction of the pulse current density $j_{pulse}$ in the center of the device. The devices are 8 terminal star-shaped devices that have four 10 μm wide arms and four 2 μm wide arms, as depicted in Fig 1 (a,b). In the first device, the wide arms are aligned parallel (0°) to the easy axis, along [1$\bar{1}$0] and [110] (Fig. 1a,c,e). In the second device, the wide arms are along [100] and [010], at an angle of 45° to the easy axis (Fig. 1b,d,f). We simulate the current-induced heat and resulting strain for different pulse configurations in our geometries using COMSOL®[20]. For the [110] star device, we simulate a straight current pulse (1.35x10$^{12}$ A/m$^2$, 0.1 ms) along the [1$\bar{1}$0] direction (white arrow in Fig. 1c). To effectively achieve the same direction of *j* in the center of the [100] star device, we use an X-shaped current pulse (1.95x10$^{12}$ A/m$^2$, 0.1 ms) through four arms (white arrows in Fig. 1d). We plot the difference between the strain along the two easy axes $\varepsilon_{[1\bar{1}0]} - \varepsilon_{[110]}$ at the surface of the NiO layer. Thus, a positive strain difference (red) corresponds to a stronger expansion of the NiO along [1$\bar{1}$0] in contrast to [110], while a negative strain difference (blue) indicates orthogonal strain. One can see from the simulations that, even if the orientation of *j* with respect to the crystallographic axis in the center of the cross is the same in the two devices (along [1$\bar{1}$0]), the strain difference induced by the current pulses is opposite in the center of the devices.

We have performed this experiment in the lab, imaging the initial domain states for both devices shown in Fig. 1a,b and the current-induced switched states after pulsing in Fig. 1e,f. For both devices we find reproducible switching, i.e. a subsequent pulse in a direction rotated by 90 degrees switches the AFM order back to the initial state. First, we note that we observe two different final states at the center of the devices depending on the strain direction and not on the direction of *j*. This is consistent with a thermomagnetoelastic mechanism that leads to opposite final states based on the strain. Second, one can see in Fig. 1f, that a single X-shaped current pulse switches the AFM domains in the arms of the cross in opposite directions with respect to the center, regardless of the direction of *j* in the arms. This can only be explained by differences in strain between the center and the region in the arms and not by a SOT dominated mechanism, which would instead induce switching towards a final state depending on the direction of *j*.

We can use the circular shape of the strain profile in the center of the device in Fig. 1d as the basis for simulations of the strain-induced magnetic switching and reproduce the observed switching through simulations (See supplementary S.1), showing that the observed switching can indeed originate from the differences in the strain profile.

Combining our results shows that the final state in our NiO/Pt samples is determined by the thermally induced strain.

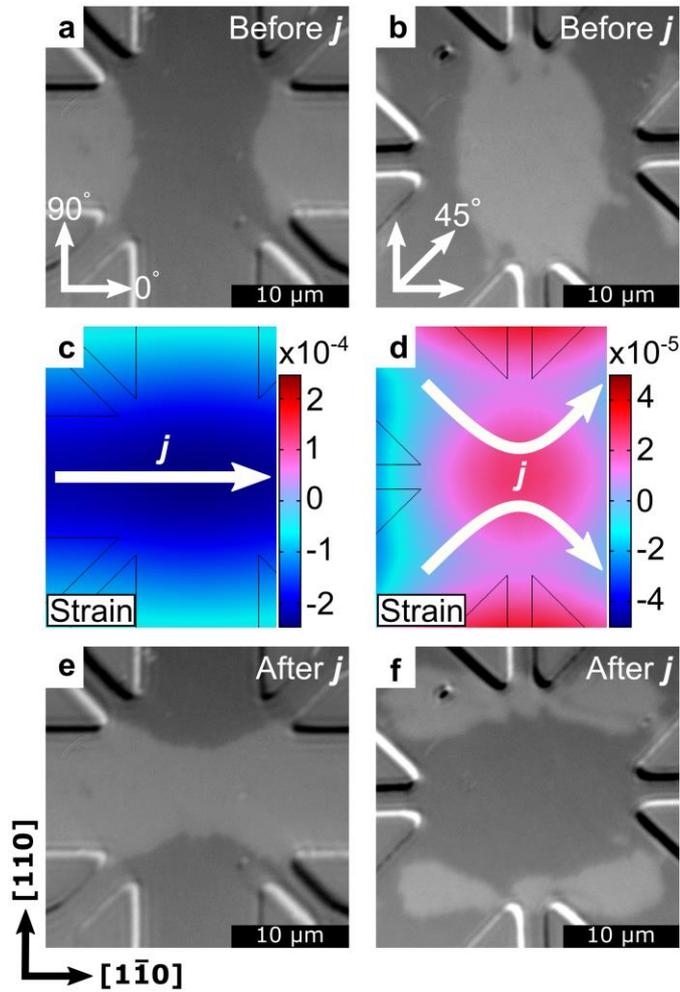

**Fig. 1:** Comparison of the switching between different device geometries. Initial domain structure in the [110] star device (a) and [100] star device (b). Simulations of the current induced strain $\varepsilon$ differences between the easy axes ($\varepsilon_{[1\bar{1}0]} - \varepsilon_{[110]}$) for a straight pulse along [1$\bar{1}$0] in the [110] star device (c) and for X-shaped pulses along [1$\bar{1}$0] in the [100] star device (d). Domain structure in the [110] star device (e) and [100] star device (f) after applying pulses in the directions indicated in (c) and (d).

**Role of temperature.** Next, we investigate whether strain alone is sufficient to switch the orientation of the Néel order and the role that current and temperature play. We consider a [100] star device and use only two arms to send a current pulse ($0.925 \times 10^{12}$ A/m$^2$, 0.1 ms) at a right angle across our device, as it is depicted in Fig. 2a (white arrow). From our simulation of the induced strains (Fig. 2a), we would expect to observe switching in the area of the current pulse and an opposite switching final state on the other side of the device. However, if we compare the imaged initial state in Fig. 2b with the final state in Fig. 2d we can only detect a switching in the wide arms, where the current flows and thus the temperature is increased due to the higher current density (see Fig. 2c). The simulated strain on the opposite side of the device is of similar magnitude as in the switched region (see Fig. 2a), however, no switching is observed. We conclude that, in this case, strain alone is not sufficient to reorient the Néel vector, and that an additional contribution by the current flow is needed either generating SOTs or heat. However, we cannot disentangle purely temperature related effects from other possible influences of the current pulse (such as SOTs) from this device geometry, due to their similar spatial distribution.

To make this distinction, we designed a cross-shaped device with a central dot, that is electrically insulated from the arms where the current flows (Fig. 3). The dot is subject to high heat and strain contributions, but it is not influenced by current-induced SOTs. The insulating gap from the current carrying electrodes is 2 µm and thus no current can flow through the NiO/Pt dot. The NiO layer below the inner circle is still affected by the electrically-induced temperature and thermomagnetoelastic strain changes, as shown by simulations of this device for a current pulse with a current density of $1.7 \times 10^{12}$ A/m$^2$ and 0.1 ms long along the [1$\bar{1}$0] direction, as depicted in Fig. 3a,c. The temperature and strain changes at the center are of similar magnitude as in the previous devices.

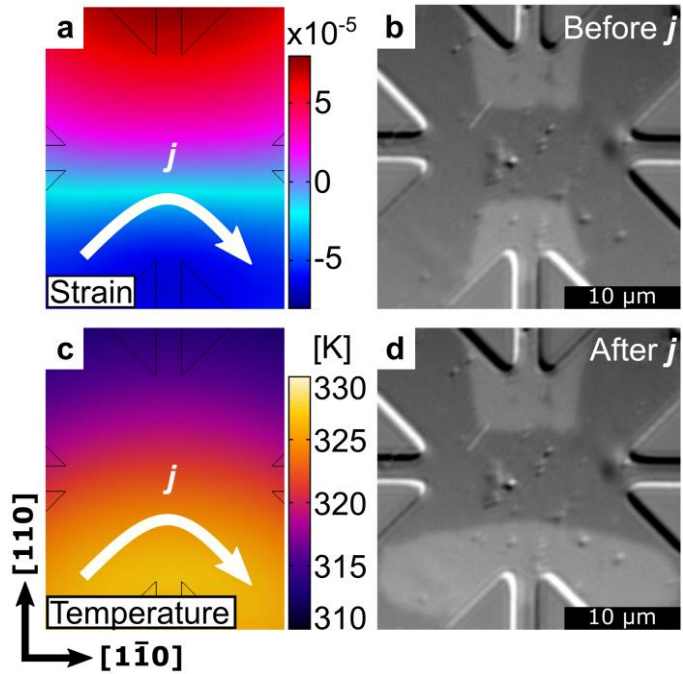

**Fig. 2:** Switching of a star-shaped device with a right angle pulse. Simulations of the current-induced strain differences between the easy axes ($\varepsilon_{[1\bar{1}0]} - \varepsilon_{[110]}$) for an edge pulse along [1$\bar{1}$0] in the [100] star device (a) and the corresponding heat profile (c). The domain structure was imaged before (b) and after the pulse application (d).

First, we image the initial domains depicted in Fig. 3b and then pulse with a current density of 1.7 x$10^{12}$ A/m². This pulse is sufficient to fully switch the device including the NiO in the region below the center dot, as can be seen in Fig. 3d. Secondly, we probe the reversibility of the switching by pulsing along the [110] direction. This change in pulsing direction results in a strain difference opposite to the first pulse (see Supplementary Fig. S.2). Therefore, the domains, as shown in Fig. 3f, are switched back to their initial state. We can observe a homogeneous switching at the center of our device, including the electrically insulated dot in the center of the cross, which indicates clearly that the current itself is not necessary to achieve switching. Moreover, in the case of a current-assisted switching mechanism, we would expect an inhomogeneous switching pattern around the etched ring due to the inhomogeneous current distribution shown in Fig. 3e, that we do not see here. Therefore, we conclude that the observed switching is enabled by the combination of heating and strain, where strain breaks the degeneracy and defines the final state, while heating is necessary to assist the switching and overcome for instance the anisotropy barrier.

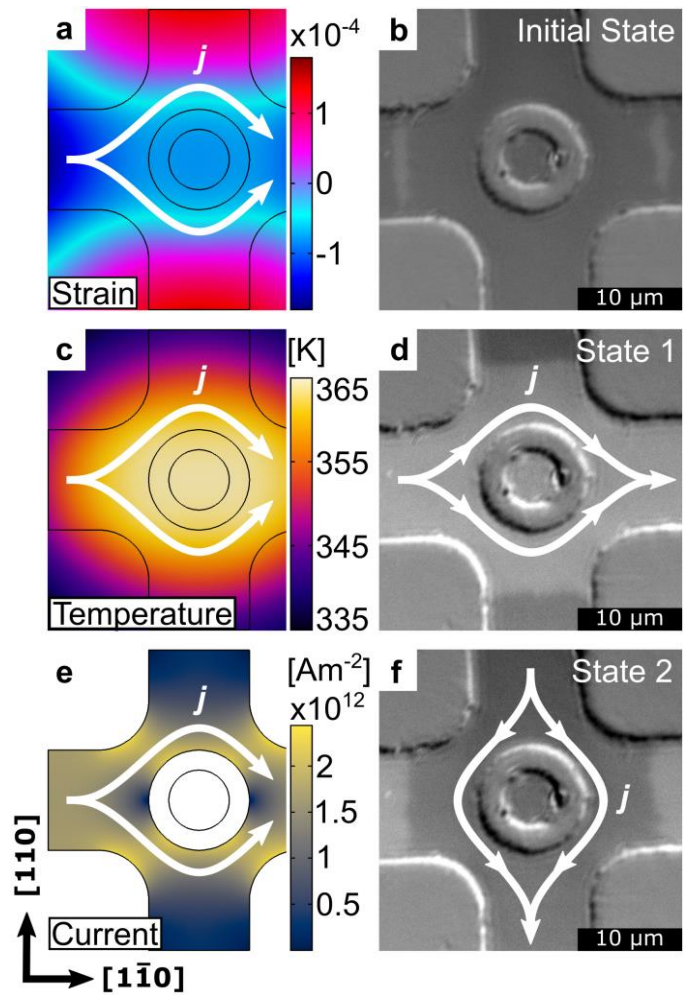

**Fig. 3:** Switching of a device with an electrically isolated area. (a) Simulations of the current-induced strain differences between the easy axis ($\varepsilon_{[1\bar{1}0]} - \varepsilon_{[110]}$) for a straight pulse along [1$\bar{1}$0] in the cross shaped device with an inner Pt circle. The corresponding temperature and current profiles are depicted in (c) and (e). The domain structure of the device was imaged before the pulse application along [1$\bar{1}$0] (b) and after (d). Domain structure after a second pulse along [110] (f).

**Electrical readout of the switching.** In addition to imaging, we conducted SMR measurements to read out the orientation of the Néel vector in the different star-shape device pulsing geometries (see Methods). Our results are consistent with the imaging and explain why, in the existing literature, different switching mechanisms were proposed based on electrical SMR measurements. We apply the pulses in the wide arms (10 µm) and read out the transverse resistance in the arms along the [010] direction, the contact scheme is depicted in Fig. 4a. Since the direction of the induced strain at the center of star-shaped devices is comparable to cross-shaped devices (see Supplementary S.3), we can compare the different switching behaviors to previous reports on crosses. To single-out the resistance changes of magnetic origin from the total resistance variation featuring also contributions of non-magnetic effects, we used the subtraction procedure described in Ref.[10]. We use series of 5 pulses each, first along the [110] direction and then along the [1$\bar{1}$0] direction. In the case of straight pulses in a [110] cross geometry (Fig. 4b, in orange) the transverse resistance (along [010]) increases after pulsing along [110] and decreases after pulses along [1$\bar{1}$0]. This indicates a final state of the Néel vector parallel to the current direction (*j* ∥ *n*), consistent with the report by Chen et al. for NiO/Pt grown on STO[14]. The same switching behavior can be observed, with lower magnitude, for the switching with right angle pulses in [100] crosses (Fig. 4b, in purple), as we previously reported in NiO/Pt Hall cross devices grown on MgO[9]. Moreover, as expected from the data shown in Fig. 1, we observe compared to the straight pulse (Fig. 4a,b in orange) an opposite final state of the switching for X shaped pulses (Fig. 4a,b in blue), where the resistance decreases after applying pulses along the [110] direction and increases after applying pulses along [1$\bar{1}$0] direction. This data indicates a final state (*j* ⊥ *n*), in line with the report of Moriyama et al. for Pt/NiO/Pt trilayers[15]. We note that different orientations of the final states with respect to the current direction in different devices have led to the proposal of several different mechanisms in literature, which is necessary if one assumes that a SOT-based mechanism dominates. By taking into account a thermomagnetoelastic switching mechanism, we can now reconcile these previous findings.

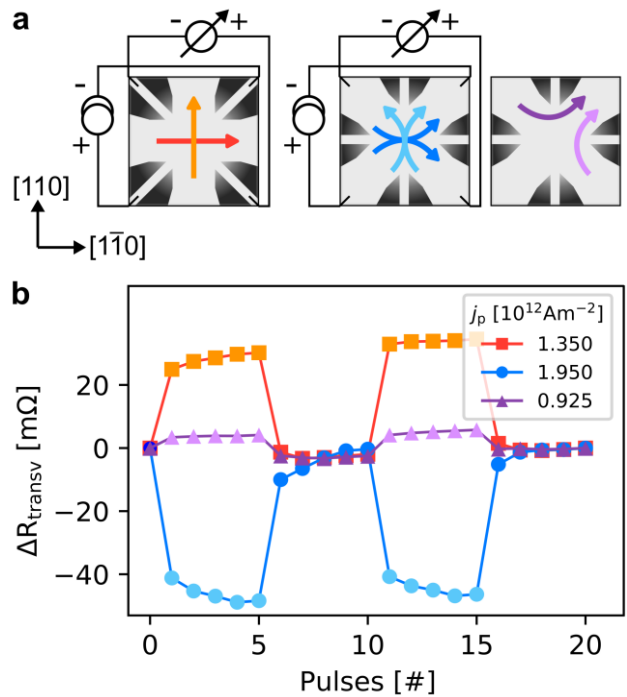

**Fig. 4:** Comparison of electrical measurements in different device and pulsing configurations. (a) Device geometries and measurement scheme with respect to the crystallographic axis. The arrows indicate the current pulse directions with light and dark color for effective [110] and [1$\bar{1}$0] pulses, respectively. (b) Corresponding electrical measurements for alternating pulse directions after subtraction of a nonmagnetic component. The magnetic origin of the shown electrical signal is confirmed by concurrent optical imaging of the domain switching.

**Discussion:**

Based on four different device and pulsing geometries, we identify the thermomagnetoelastic mechanism in combination with thermal excitations to be responsible for the observed current-induced switching in our NiO/Pt bilayers and show how previously observed contradicting reports on the final switching state of NiO/Pt can be unified by considering a thermomagnetoelastic switching mechanism and different geometries.

With these robust findings, we suggest that the thermomagnetoelastic switching mechanism is not solely limited to NiO thin films and can lead to switching in other materials and systems, especially where the switching of thick layers was observed. There is furthermore evidence that the thermomagnetoelastic mechanism can dominate over SOTs also in other AFMs with large magnetoelastic coupling such as Hematite[8] and CoO[13] and might also contribute to some recent data obtained in ferrimagnetic insulators[21]. As a key point, we find that additionally the heating of the AFM is an essential component to enable the thermomagnetoelastic AFM switching and thus not only the strain but also the temperature profile has to be considered. In addition, we note that the final state of the observed switching in NiO/Pt with straight pulses (*n* ∥ *j*) is opposite to the recently reported switching in CoO/Pt (*n* ⊥ *j*)[13]. These differences between materials can be traced back to an opposite sign of the magnetoelastic coupling, resulting in different preferred orientations of the Néel vector[13]. Future investigations on current-induced switching of the Néel vector have to take these findings into account in order to differentiate between switching due to SOTs or thermomagnetoelastic effects in heated areas. While multiple mechanism can coexist, SOT based switching might dominate in thinner layers or materials with lower magnetostriction. Our data sheds light on the switching observed in insulating AFMs and which mechanism is responsible for the switching observed in these materials, being important not only to understand the phenomena, but also in view of possible applications.

Here, we identified that strain alone is not sufficient to switch our AFM thin film, but heat is additionally required. Since both, strain and heat can be generated in a non-contact manner, this switching mechanism potentially enables a new class of non-contact writing schemes in which high current densities flowing through the device leading to device degradation due to electromigration are absent. For instance, one can apply global strain externally, e.g. by strain gauges or piezoelectric materials, and then use ultra-fast concurrent local heating by a focused laser beam to rapidly switch AFM elements. Such an approach drastically simplifies the patterning process and increases the surface of the device available for information storage. In combination with non-contact read outs by the magnetooptical Kerr effect or coupled ferromagnetic layers, electrical contacts and adjacent HM layers are no longer required to read and write the insulating AFM domains. Such non-contact geometries can increase the efficiency and reliability of a device, since losses due to resistance and subsequent heating in unwanted areas are avoided. Overall, the discovery of a thermally assisted thermomagnetoelastic switching mechanism establishes an alternative, possibly non-contact writing scheme for insulating AFMs which can pave the way for alternative functionalization of this class of materials.

**Methods section:**

**Samples and Devices.** The NiO(001)(10nm)/Pt(2nm) bilayers that are presented here were grown epitaxially on MgO(100) substrates by reactive sputtering from a Ni target at high temperature and DC sputtering of the Pt layer at room temperature. The growth parameters are reported elsewhere (in Ref.[10]). The antiferromagnetic ordering of the NiO was verified by x-ray magnetic linear dichroism (XMLD) and the absence of x-ray magnetic circular dichroism (XMCD). Due to the lattice mismatch between NiO (4.18 Å) and MgO (4.21 Å) the samples were grown under tensile strain. Electron beam lithography and subsequent ion beam etching of the Pt layer were used to fabricate the different devices.

**Writing the AFM state.** For the writing of the AFM state we used the same technique described in Ref.[10], employing a Keithley 6221. We applied a single 0.1 ms long electrical pulse to switch the different devices. Depending on the device and pulsing geometry the resistances of our devices varied, and we chose different current densities to observe a reversible switching of large domains at the center of the device. Reversible switching was achieved in all cases by rotating the effective current direction of the pulse geometry by 90 degrees.

**Imaging and reading out the AFM state.** We employ a magneto-optical imaging technique of the antiferromagnetic domains, based on a commercial Kerr microscope, as described in Ref.[10,18,22]. We image the AFM domains before and after switching in different devices with different pulsing geometries. The imaging and electrical measurements have been performed with the same setup and configuration described in Ref.[10] using a Keithley 2400 as a source of the measurement current and a Keithley 2182A nanovoltmeter for the read-out. The routing between the different reading and pulsing configurations is done electronically by a matrix switch.

**Simulations.** We utilize (3D) finite element modeling in COMSOL Multiphysics® to study the current-induced heating and subsequent strain contributions in our different device and pulsing geometries. We apply 0.1 ms pulses with varying current densities across the platinum layers. The device and substrate dimensions correspond to the real experimental environment, the contact pads are simplified to reduce computational efforts. We modeled the end of the current pulse with a step function and displayed here the strain and heat distributions at the surface of the NiO layer after 0.1 ms. More details on the parameters used can be found in the supplementary.

**Acknowledgements**

The authors thank T. Reimer for skillful technical assistance. L.B acknowledges the European Union's Horizon 2020 research and innovation program under the Marie Skłodowska-Curie grant agreements ARTES number 793159. L.B. and M.K. acknowledge support from the Graduate School of Excellence Materials Science in Mainz (MAINZ) DFG 266, the DAAD (Spintronics network, Project No. 57334897) and all groups from Mainz acknowledge that this work was funded by the Deutsche Forschungsgemeinschaft (DFG, German Research Foundation) - TRR 173 – 268565370 (projects A01, A03, A11, B02, and B12). M.K. acknowledge financial support from the Horizon 2020 Framework Programme of the European Commission under FET-Open grant agreement no. 863155 (s- Nebula). This work was also supported by ERATO "Spin Quantum Rectification Project" (Grant No. JPMJER1402) and the Grant-in-Aid for Scientific Research on Innovative Area, "Nano Spin Conversion Science" (Grant No. JP26103005), Grant-in-Aid for Scientific Research (C) (Grant No. JP20K05297) from JSPS KAKENHI, Japan.

COMSOL® and COMSOL Multiphysics® are registered trademarks of COMSOL AB.